\documentclass[11pt,a4paper]{article}
\usepackage[utf8]{inputenc}
\usepackage[english]{babel}
\usepackage{amsmath}
\usepackage{amsfonts}
\usepackage{amssymb}
\usepackage{amsthm}
\usepackage{bm}

\newtheorem{theorem}{Theorem}
\newtheorem{definition}{Definition}

 \def\be{\begin{equation}}
\def\ee{\end{equation}}
 \def\ba{\begin{align}}
\def\ea{\end{align}}
\def\bea{\begin{eqnarray}}
\def\eea{\end{eqnarray}}

\def\m{\mu}
\def\n{\nu}

\def\r{\rho}
\def\s{\sigma}

\def\D{\nabla}

\def\tD{\tilde{\nabla}}

\begin{document}
%%%%%%%%%%%%%%%%%%%%%%%%%%%%%%%%
%%%%%%%%%%%%%%%%%%%%%%%%%%%%%%%%
%%%%%%%%%%%%%%%%%%%%%%%%%%%%%%%%
%%%%%%%%%%%%%%%%%%%%%%%%%%%%%%%%
%%%%%%%%%%%%%%%%%%%%%%%%%%%%%%%%
\clearpage
\begin{titlepage}
\title{\vspace{-2cm} 
\begin{flushright}
{\normalsize IFT-UAM/CSIC-16-018},
{\normalsize FTUAM-16-5}
\end{flushright}
\vspace{0.5cm} 
{\bf Anomalies, equivalence and renormalization of cosmological frames}}

\author{Mario Herrero-Valea\\[2mm]
{\small\it Instituto de F\' isica Te\'orica UAM/CSIC,}\\[-1mm]
{\small\it C/ Nicolas Cabrera, 13-15, 
C.U. Cantoblanco, 28049 Madrid, Spain}\\
{\small \it and}\\
{\small\it Departamento\! de\! F\'isica\! Te\'orica,}\\
{\small \it \! Universidad\! Aut\'onoma\! de\! Madrid,\! 20849\! Madrid,\! Spain}
}
\date{}
\maketitle

\begin{abstract}
We study the question of whether two frames of a given physical theory are equivalent or not in the presence of quantum corrections. By using field theory arguments we claim that equivalence is broken in the presence of anomalous symmetries in one of the frames. This is particularized to the case of the relation between the Einstein and Jordan frames in scalar-tensor theories used to describe early Universe dynamics. Although in this case a regularization that cancels the anomaly exists, the renormalized theory always develop a non-vanishing contribution to the S-matrix that is present only in the Jordan frame, promoting the different frames to different physical theories that must be UV completed in a different way.
\end{abstract}
\end{titlepage}
%%%%%%%%%%%%%%%%%%%%%%%%%%%%%%%
\tableofcontents

%%%%%%%%%%%%%%%%%%%%%%%%%%%%%%%%
%%%%%%%%%%%%%%%%%%%%%%%%%%%%%%%%
%%%%%%%%%%%%%%%%%%%%%%%%%%%%%%%%
%%%%%%%%%%%%%%%%%%%%%%%%%%%%%%%%
%%%%%%%%%%%%%%%%%%%%%%%%%%%%%%%%
%%%%%%%%%%%%%%%%%%%%%%%%%%%%%%%%
\section{Introduction}
Physics is a universal way of describing phenomena in our Universe. In particular, it is a field that is able to make predictions about events to come. For this property to be satisfied, the main requirement for any physical theory is that these predictions cannot depend on the observer who makes them neither on the ruler she uses. Indeed, Galilean Relativity as well as the Special and General theories of Relativity contain this assumption deep in their core. 

There is some tension however when Quantum Field Theory enters into game. It has been known for some time\cite{Vilkovisky:1984st} that the generator of 1PI diagrams, the quantum effective action, is not invariant under field reparametrizations and even its invariance under internal gauge symmetries has generated some controversy\cite{Grisaru:1975ei,Kallosh:1974yh,DeWitt:1965jb}. 

In the recent years, observations of experiments like the WMAP or Planck satellites\cite{Ade:2015lrj} have opened a window to study cosmological phenomenon which occur close to the Planck scale, where the first corrections to classical observables coming from quantum gravitational dynamics may play an important role. The most famous example of this is inflation, where a huge number of different models have been proposed to make the Universe expand exponentially in its very first few moments\cite{Guth:1980zm,Linde:1981mu,Linde:1983gd}. Some of these models rely on the existence of a new scalar degree of freedom, the inflaton field\footnote{In some models, the Standard Model Higgs field takes the role of the inflaton.}, non-minimally coupled to gravity through a $\phi^2 R$ term\cite{Bezrukov:2007ep,Kallosh:2013wya,Kallosh:2013hoa, Roest:2013fha}. The complications that this coupling introduces when trying to compute quantum corrections are huge and then it is common to transform the field variables, the metric and the inflaton, to go from this so called Jordan frame to the Einstein frame, where the non-minimal coupling is not present and more standard intuition can be used. Computing quantum corrections in the Einstein frame is easier and afterwards they can be taken back to the original frame by reverting the field redefinition.

Many concerns have appeared in the last years when dealing with this situation (cf. \cite{Kamenshchik:2014waa} and references therein). In principle, the already mentioned non-equivalence of the quantum effective action could be fatal if one wants to extract physical predictions from this setting. However, it can be proven both in general and for particular actions\cite{Kamenshchik:2014waa, Alvarez:2014qca}, that the S-matrix computed in both frames is indeed equivalent, provided that the fields are transformed accordingly, this saving the day and the equivalence. Nonetheless, there is a caveat in all these arguments. If an anomalous symmetry is present in one of the frames, then the equivalence is broken precisely by the anomalous contributions and one can find at least an observable, corresponding to the conserved current of the corresponding symmetry, which does not transform appropriately, giving inequivalent physics in both frames.

In this work we unveil the origin and consequences of this \emph{anomalous equivalence of frames}, first by general arguments based on the structure of the quantum effective action and afterwards with an explicit example of a scale invariant scalar-tensor theory. Our conclusion, however, is not particular for this model and can in principle be extended to more general QFT's, the only hypothesis being the exchange of an exact symmetry by an anomalous one due to the field transformations.

%%%%%%%%%%%%%%%%%%%%%%%%%%%%%%%%
%%%%%%%%%%%%%%%%%%%%%%%%%%%%%%%%
%%%%%%%%%%%%%%%%%%%%%%%%%%%%%%%%
%%%%%%%%%%%%%%%%%%%%%%%%%%%%%%%%
%%%%%%%%%%%%%%%%%%%%%%%%%%%%%%%%
%%%%%%%%%%%%%%%%%%%%%%%%%%%%%%%%
\section{Anomalous equivalence of frames}
One of the main assumptions of physics is that predictions cannot depend on the observer who makes an experiment neither on the variables she uses to describe it. Unifying both aspects, it is assumed that physical phenomena cannot depend on the choice of \emph{frame}. This is the minimal requirement one can ask to a well-behaved physical theory. Indeed, classical physics satisfies this in a very simple way, quantities in one frame are just the transformation of quantities in an equivalent frame, being Lorentz transformations a well-known example of this kind of situation. One can revert this argument and use it to actually define what equivalence between frames means
\begin{definition}[Equivalence between frames]\label{th:equivalence}
Two frames related by a transformation of the field variables of the form $Q'=Q'(Q)$ are equivalent if any physical quantity $A(Q)$ satisfies
\begin{align}
A'(Q')=A(Q)
\end{align}
\end{definition}

In particular, this applies to the action $S'(Q')=S(Q)$, ensuring that stationary trajectories in one frame correspond also to stationary trajectories in any other. The full equivalence of classical observables then follows trivially from this fact.

The situation is not so easy in Quantum Field Theory(QFT). Indeed, it was noted quite some time ago\cite{Vilkovisky:1984st} that one of the main objects in QFT, the quantum effective action $\Gamma[\mathbf{Q}]$, was not invariant under a change of frames. In particular, let $S[Q]$ be a classical action depending on a set of fields $Q$, then we can define the effective potential $W[J]$ through a path integral with sources $J$
\begin{align}
e^{i W[J]}=\int [{\cal D} Q]\;e^{iS[Q]+i\int d^4x\; J\cdot Q}
\end{align}
and the quantum effective action as a Legendre transform of this
\begin{align}
\Gamma[\mathbf{Q}]=W[J]-\int d^4x\; J \cdot\mathbf{Q}
\end{align}
where $\mathbf{Q}$ is the mean field
\begin{align}
\mathbf{Q}=\frac{\delta W}{\delta J}
\end{align}
which solves the effective equations of motion given by the quantum effective action
\begin{align}
\frac{\partial \Gamma}{\partial \mathbf{Q}}=-J
\end{align}

At the one-loop level, $\Gamma[\mathbf{Q}]$ can be exactly integrated to yield
\begin{align}\label{eq:one-loop_Gamma}
\Gamma[\mathbf{Q}]=S[\mathbf{Q}]+\frac{i}{2}\log\left[\text{Det}\left(\frac{\partial^2 S}{\partial Q^2}\right)\right]
\end{align}
which under a change of frames $Q'=Q'(Q)$ transforms as
\begin{align}\label{eq:transformation_Gamma}
\Gamma'[\mathbf{Q}']=S'[\mathbf{Q}']+\frac{i}{2}\log\left[\text{Det}\left(\frac{\partial^2 S'}{\partial Q'^2}+\left(\frac{\partial Q'}{\partial Q}\right)^2 \frac{\partial^2 Q}{\partial Q'^2}\frac{\partial S'}{\partial Q'}\right)\right]
\end{align}

Therefore, only quantities which are evaluated on-shell, a requirement that almost uniquely selects the S-matrix, satisfy equivalence in the sense of Definition \ref{th:equivalence}. In particular, \cite{Vilkovisky:1984st} pointed out that other objects, like for instance the mean field obtained by differentiating $\Gamma[\mathbf{Q}]$, would be affected by this difference, compromising equivalence. They tracked the reason for this disagreement to the fact that $\Gamma[\mathbf{Q}]$, as it is defined, is not a scalar quantity on the manifold spanned by the field configurations of $Q$, so any redefinition of these will change its value. The obvious solution was then to covariantize the quantum effective action, arriving to what we name \emph{Unique Effective Action.} The need of using this redefinition or not has been discussed before in the literature\cite{Calcagni:2015vja,Falls:2015qga, Calcagni:2014xca,Steinwachs:2013tr}, but we do not aim to enter into this dilemma in the present work. Here, instead, we want to unveil another different issue that jeopardizes equivalence.

Although \eqref{eq:one-loop_Gamma} and \eqref{eq:transformation_Gamma} are completely general formulas at the one-loop level, there is a caveat in their derivation. If the variables of one frame, lets say $Q'$, realize different symmetries than the ones in the other frame and one of those happens to be anomalous, then equivalence is compromised.

Let us again be explicit and assume that the action $S'[Q']$ is invariant under some transformation given by
\begin{align}
Q'\rightarrow Q'+\delta_\epsilon Q'+O(\epsilon^2)
\end{align}
where $\epsilon$ is the infinitesimal generator of the transformation. 

Then, classically, Noether's theorem implies
\begin{align}
\frac{\delta S'}{\delta Q'}\delta_\epsilon Q'=\D'_\mu J'^\mu
\end{align}

In the case of a Quantum Field Theory, this translates into the corresponding Ward-Takahashi identity
\begin{align}\label{eq:ward_Q'}
\bigg\langle \frac{\delta \Gamma'}{\delta \mathbf{Q}'}\delta_\epsilon \mathbf{Q}'\bigg\rangle= \langle \D'_\mu J'^\mu\rangle
\end{align}
which must hold when quantum corrections are considered unless we find an anomaly. In that case, they will be modified by new terms in the rhs.

In the $Q$ frame, if equivalence as given by Definition \ref{th:equivalence} holds, the transformed version of the identity must be still true as an operator equation
\begin{align}\label{eq:ward_Q}
\bigg\langle \frac{\delta \Gamma}{\delta \mathbf{Q}}\delta_\epsilon \mathbf{Q}\bigg\rangle= \langle \D_\mu J^\mu\rangle
\end{align}
as a consequence of a different symmetry given by
\begin{align}
Q\rightarrow Q+\delta_\epsilon Q=Q+\frac{\partial Q}{\partial Q'}\delta_\epsilon Q'
\end{align}

Remarkably, the concrete realization of the symmetries in the different frames will be as well different if the transformations are non-linear. In particular, it could happen that the symmetry in the $Q$ frame is exact\footnote{By exact we mean that $\delta_{\epsilon}Q$ gets no anomalous contributions from quantum corrections.} and therefore the condition \eqref{eq:ward_Q} will hold at any order in the loop expansion. But, if the corresponding symmetry is anomalous in the $Q'$ frame, this means that the transformed version \eqref{eq:ward_Q'} of the identity will not hold but instead it will get new contributions. We find then a source of inequivalence that can be sharply stated 
\begin{theorem}[Anomalous equivalence]\label{th:inequivalence}
If the frame $Q'$ related to the frame $Q$ as $Q'=Q'(Q)$ is invariant under an anomalous symmetry $Q'\rightarrow Q'+\delta_\epsilon Q'$, which corresponds to an exact symmetry $Q\rightarrow Q+\delta_\epsilon Q$ in the frame $Q$, equivalence of frames does not hold.
\end{theorem}
%%%%%%%%%%%%%%%%%%%%%%%%%%%%%%%%%
%%%%%%%%%%%%%%%%%%%%%%%%%%%%%%%%
\section{Inequivalence of cosmological frames}
Let us now particularize our result to the case of interest of cosmological frames. For that, we start by introducing a scalar-tensor model in the Jordan Frame written in euclidean signature
\begin{align}
S_{J}[g_{\m\n},\phi]=\int d^4 x \sqrt{|g|}\;\left(-\xi \phi^2 R+\frac{1}{2}\nabla_\mu \phi \nabla^\mu \phi +\lambda \phi^4 \right)
\end{align}
where $\phi$ is a scalar field and $\lambda$ and $\xi$ are dimensionless couplings. As it stands, this action is invariant under the scale transformations
\begin{align}\label{eq:scale_transformations}
g_{\m\n}\rightarrow \Omega^2 g_{\m\n},\qquad \phi\rightarrow \Omega^{-1} \phi
\end{align}

This will be our frame $Q'$ and the fact that $S_J[g_{\m\n},\phi]$ is scale-invariant is crucial for our result. It will actually be scale invariance what takes the role of the anomalous symmetry in this frame. We consider this action exclusively when $\phi\neq 0$. In other case, perturbation theory would be broken by the presence of a strong coupled gravitational sector\cite{tHooft:2011aa}.

We now define the other frame $Q$ as the corresponding Einstein frame of this theory. It is obtained by doing the following transformations
\begin{align}\label{eq:field_redefinition}
\tilde{g}_{\m\n}= \frac{\xi \phi^2}{M_p^2}\; g_{\m\n},\qquad \varphi=M_p \sqrt{\frac{1}{\xi}-12}\;\log(\phi)
\end{align}
where $M_p$ is the Planck mass\footnote{In the case of $\xi=\frac{1}{12}$ the symmetry is enhanced to Weyl invariance and there is no scalar field in the Einstein frame. This is the case examined in \cite{Alvarez:2014qca}.}. Thus, we arrive to the action
\begin{align}\label{eq:Einstein_action}
S_E[\tilde{g}_{\m\n},\varphi]=\int d^4x \sqrt{|\tilde{g}|}\;\left(  -M_p^2 \tilde{R} +\frac{1}{2}\tD_\m \varphi \tD^\m \varphi +\frac{\lambda M_p^4}{\xi^2}\right)
\end{align}

It is worth to point here that these actions are non-renormalizable upon the inclusion of dynamical gravity. Thus, they must be understood as the lowest order terms in an expansion in inverse powers of some ultraviolet scale. Higher order contributions will be produced by quantum loops and they will be related by the field transformations as well\cite{Kallosh:1974yh,DeWitt:1965jb}. We will come back to this point later.

In the Einstein frame we do not have any scale invariance any more because the Einstein-Hilbert term $M_p^2 \tilde{R}$ breaks it explicitly through the presence of the Planck mass. It is substituted by a shift symmetry in the scalar field
\begin{align}
\varphi\rightarrow \varphi + \tilde{\Omega}
\end{align}

Then, at the classical level and at tree-level in the quantum theory, the transformed version of the scale invariant Ward identity must still be true at the light of shift invariance. Indeed, such Ward identity is, in the Jordan frame, just the statement that the trace of the energy-momentum tensor is a total derivative on the mass-shell
\begin{align}
\bigg \langle \;T_{\phantom{q}\m}^{\m}-\phi\frac{\delta S_J}{\delta \phi} \bigg\rangle =\langle\nabla_\m J^\m\rangle
\end{align}

However, in the presence of gravity, what we would define as the energy-momentum tensor\footnote{By energy-momentum tensor we mean the \emph{Belinfante-Rosenfeld energy-momentum tensor}\cite{BELINFANTE1940449}, which corresponds to the source of the gravitational field and is obtained by functional differentiating the lagrangian density
\begin{align}
T^{\m\n}=\frac{2}{\sqrt{|g|}}\frac{\delta (\sqrt{|g|}\;{\cal L})}{\delta g_{\m\n}}
\end{align}} is no more than the gravitational equation of motion. Thus, scale invariance forces its trace to be a total derivative
\begin{align}
\bigg\langle 2g_{\m\n}\frac{\delta S_J}{\delta g_{\m\n}}-\phi\frac{\delta S_J}{\delta \phi}\bigg\rangle=\langle \nabla_\m J_J^\m\rangle
\end{align}
where the current is simply\footnote{Note that, as expected, in the Weyl invariant point $\xi=\frac{1}{12}$ this vanishes identically.}
\begin{align}
J_J^\m=(1-12\xi)\;\phi \D^\m\phi
\end{align}

If we now go to the Einstein frame, the aforementioned shift symmetry will ensure that we have the equivalence relation
\begin{align}
\bigg\langle 2g_{\m\n}\frac{\delta S_J}{\delta g_{\m\n}}-\phi\frac{\delta S_J}{\delta \phi}\bigg\rangle=\langle \nabla_\m J_J^\m\rangle  \xrightarrow{\phantom{aaa}Q'\rightarrow Q  \phantom{aaa}}   \bigg\langle 2\tilde{g}_{\m\n}\frac{\delta S_E}{\delta \tilde{g}_{\m\n}}-\varphi\frac{\delta S_E}{\delta \varphi}\bigg\rangle=\langle \tilde{\nabla}_\m J_E^\m\rangle
\end{align}
which completely holds at tree-level.

Let us consider now quantum corrections to this situation. If these are non-vanishing, their effect will be to modify the effective equations of motion that enter into the Ward identity by substituting the classical action $S_i$ by the quantum effective action $\Gamma_i$ and the classical fields by their mean values so that the previous relation changes to
\begin{align}
\bigg\langle 2\mathbf{g}_{\m\n}\frac{\delta \Gamma_J}{\delta \mathbf{g}_{\m\n}}-\bm{\phi}\frac{\delta \Gamma_J}{\delta \bm{\phi}}\bigg\rangle=\langle \nabla_\m J_J^\m\rangle  \xrightarrow{\phantom{aaa}Q'\rightarrow Q  \phantom{aaa}}   \bigg\langle 2\mathbf{\tilde{g}}_{\m\n}\frac{\delta \Gamma_E}{\delta \mathbf{\tilde{g}}_{\m\n}}-\bm{\varphi}\frac{\delta \Gamma_E}{\delta \bm{\varphi}}\bigg\rangle=\langle \tilde{\nabla}_\m J_E^\m\rangle
\end{align}

The fact that gravity is dynamical reflects here in the fact that the divergence of the current must be proportional to the equations of motion.

As it stands, the relation on the Einstein frame is still the transformation of the one in the Jordan frame. However, in that frame, this is not the real identity that is preserved. The reason is that, in general, scale invariance is anomalous due to the appearance of a non-vanishing variation of the pole term in the regularized effective action under the scale transformations \eqref{eq:scale_transformations}. Then, the corresponding Ward-Takahashi identity must be modified by the anomalous terms ${\cal A}[g_{\m\n},\phi]$
\begin{align}\label{eq:Ward_anomalous}
\bigg\langle 2\mathbf{g}_{\m\n}\frac{\delta \Gamma_J}{\delta \mathbf{g}_{\m\n}}-\bm{\phi}\frac{\delta \Gamma_J}{\delta \bm{\phi}}\bigg\rangle= {\cal A}[g_{\m\n},\phi]+\langle\D_\m J_J^\m\rangle
\end{align}

Therefore the equivalence \emph{does not hold any more.} If we wanted to preserve it, then we need to consider not only the constraints imposed by the effective equations of motion
\begin{align}
\frac{\delta \Gamma_J}{\delta \mathbf{g}_{\m\n}}=\frac{\delta \Gamma_J}{\delta \bm{\phi}}=0
\end{align}
but also the ones required by anomaly cancellation
\begin{align}
 {\cal A}[g_{\m\n},\phi]=0
\end{align}
which have no counterpart in the Einstein frame. Moreover, the presence of the anomaly triggers a scattering amplitude\footnote{The prototypical example of this effect is the $\pi^0\rightarrow \gamma\gamma$ decay in chiral theories, which is driven by the axial anomaly.}
\begin{align}
\left< \sigma \right|  {\cal A}[g_{\m\n},\phi]\left| 0\right>\neq0
\end{align}
between some state of the theory and the vacuum, which is however absent in the Einstein frame. This makes the frames, and perhaps the physical consequences derived from then, completely inequivalent.
%%%%%%%%%%%%%%%%%%%%%%%%%%%%%%%%
%%%%%%%%%%%%%%%%%%%%%%%%%%%%%%%%
\section{Scale anomalies}
Up to here we have argued that in the presence of an anomalous scale invariance, equivalence between Einstein and Jordan frames is broken. Now we take care of the explicit form of the anomalous term ${\cal A}[g_{\m\n},\phi]$. For that, let us keep in mind that we are dealing with non-renormalizable theories, so that at every loop order we will find new counterterms to be introduced in the action. Thus we can keep the computation under control at most order by order in the loop expansion. Moreover, in the Jordan frame, where we have a non-linear coupling between curvature and the scalar field $\phi$, standard power counting arguments do not hold\cite{Alvarez:2014qca} and we expect to find counterterms of the form
\begin{align}\label{eq:singular_operators}
\sqrt{|g|}\;\frac{(\D_\m \phi\D^\m\phi )^2}{\phi^4},\quad \sqrt{|g|}\;\frac{(\square \phi)^2}{\phi^2},\quad \sqrt{|g|}\;\frac{R^3}{\phi^2},\quad ...
\end{align}
which are singular in the limit $\phi\rightarrow 0$. For this particular example, the first two terms appear at one-loop while the last one corresponds to a two-loops correction.

The easiest way to compute corrections in the Jordan frame is then to exploit the pretending equivalence. That is, even if we know of a particular current which is not equivalent between the two theories (that of scale invariance), the only effect can come from anomalies and therefore, the counterterms will be still related by the frame transformation $Q'=Q'(Q)$. Indeed, that the counterterms are equivalent on-shell has been proven in many different works\cite{Alvarez:2014qca, Steinwachs:2013tr, Kamenshchik:2014waa}. Thus, we can compute the regularized quantum effective action in the Einstein frame, transform it back to Jordan frame and then obtain the anomalous terms and the anomalous Ward identity from them by renormalizing in this frame. Let us remark, however, that we are using this trick here only as a way to keep the computational complexity of this letter as low as possible. One could instead compute the corrections directly in the Jordan frame, where a complicated gauge fixing sector is required, arriving to the same results for the dimensionally regularized divergences of the theory.

In the Einstein frame, power counting arguments work and the loop expansion can be regarded as an expansion in scaling dimension. If we use dimensional regularization, then the $L$-loop level contribution to the effective action reads
\begin{align}\label{eq:counter_einstein}
\Gamma[\tilde{g}_{\m\n},\varphi]_L=\frac{1}{n-4}\int d^n x \sqrt{|g|}\; \sum_{{\cal O}} {\cal O}_L[\tilde{R}_{\m\n\r\s},\varphi]+\text{finite}
\end{align}
where the operators $ {\cal O}_L[\tilde{R}_{\m\n\r\s},\varphi]$ are scalars quantities of scaling dimension $2+2L$, with their mass dimension suppressed by powers of $M_p^2$ in order to get a four-dimensional admissible term after renormalizing. Here we have explicitly written only the pole term since the finite residues will be irrelevant for our discussion of anomalies. Although the complete list of such operators grows very quickly with $L$, those depending only on the curvature can be simply given as all the possible contractions of the Riemann tensor and derivatives with the right scaling. In the particular case of $L=1$, they are scale invariant in four dimensions and read\cite{'tHooft:1974bx,Vassilevich:2003xt}
\begin{align}
R_{\m\n\r\s}R^{\m\n\r\s},\quad R_{\m\n}R^{\m\n},\quad R^2,\quad \square R
\end{align}

Moreover, all the counterterms preserve identically the shift symmetry, since only derivatives of the scalar field can be generated by quantum loops. No anomaly in this symmetry can appear when using dimensional regularization.

In the Jordan frame, the counterterms will correspond to the transformation of \eqref{eq:counter_einstein} under the map of frames
\begin{align}
\{\tilde{g}_{\m\n},\varphi\}\longrightarrow \{g_{\m\n},\phi\}
\end{align}
giving for the effective action
\begin{align}
\Gamma[g_{\m\n},\phi]_L=\frac{1}{n-4}\int d^n x \sqrt{|g|}\; \sum_{{\cal O}'} {\cal O}'_L[R_{\m\n\r\s},\phi]+\text{finite}
\end{align}
where now the supresion of mass dimension will be done with powers of $\phi$, suppressing scaling dimension at the same time. This happens because the transformations \eqref{eq:field_redefinition} transmute the Planck mass into powers of the scalar field. Therefore we will obtain the singular operators introduced in \eqref{eq:singular_operators}. In particular, this was computed at one loop in \cite{Alvarez:2014qca} directly in the Jordan frame, obtaining
\begin{align}
\nonumber \Gamma[g_{\m\n},\phi]_1&=\frac{1}{n-4}\frac{1}{16\pi^2}\int d^n x \sqrt{|g|}\;\bigg\{\frac{71}{60}C_{\m\n\r\s}C^{\m\n\r\s}+\frac{1259}{1440}\frac{(1-12\xi)^2}{\xi^2} \frac{(\nabla \phi)^4}{\phi^4}+\\
&+\frac{1484}{1440} \frac{1-12\xi}{\xi^2}\lambda (\nabla \phi)^2-\frac{371}{180}\frac{\lambda^2}{\xi^2}\phi^4\bigg\}
\end{align}
for the on-shell divergences in dimensional regularization. Here $C_{\m\n\r\s}$ is the Weyl tensor. Note also that all the non-Weyl invariant operators disappear when $\xi=\frac{1}{12}$ as it must be.

There is a subtle point to be discussed here. In \eqref{eq:field_redefinition} we defined the transformations $Q\rightarrow Q'$ in such a way that we go from a scale invariant theory in the Jordan frame in four dimensions to the Einstein frame with Einstein-Hilbert term $-M_p^2 R$. However, this is only true in four dimensions and since we are dealing here with dimensional regularization, it is reasonable to ask whether we should stick to this definition or upgrade it to a dimension dependent one, meaning that we redefine the fields as
\begin{align}\label{eq:field_redefinition_n}
\tilde{g}_{\m\n}=M_p^{-2}\xi^{\frac{2}{n-2}} \phi^\frac{4}{n-2}\; g_{\m\n},\qquad \varphi=M_p \sqrt{\frac{1}{\xi}-12}\;\log(\phi)
\end{align}
in such a way that the bare action\footnote{Mind the power of $\phi$ in the potential term.}
\begin{align}
S_{J}[g_{\m\n},\phi]=\int d^n x \sqrt{|g|}\;\left(-\xi \phi^2 R+\frac{1}{2}\nabla_\mu \phi \nabla^\mu \phi +\lambda \phi^\frac{2n}{n-2} \right)
\end{align}
goes now exactly to the Einstein frame action \eqref{eq:Einstein_action} in any space-time dimension $n$. Moreover, starting from \eqref{eq:Einstein_action}, only using the inverse of \eqref{eq:field_redefinition_n} we get a scale invariant action in any dimension.

This is relevant for the discussion here because if we stick to the original definition \eqref{eq:field_redefinition}, we find that the counterterms in the Jordan frame only coincide with those computed directly there if we perform the transformation in $n=4$, and these are invariant under scale transformations in, \emph{and only in} four dimensions. This means that, since we are defining our counterterms in dimensional regularization, they will transform at the infinitesimal level as
\begin{align}
\delta \left(\sqrt{|g|}\;{\cal O}'_L[R_{\m\n\r\s},\phi]\right)=(n-4)\;\omega \; \sqrt{|g|}\;{\cal O}'_L[R_{\m\n\r\s},\phi]
\end{align}
where $\Omega=1+\omega+O(\omega^2)$.

Therefore, in the limit $n\rightarrow 4$ in which we remove the regularization, we find a finite residue which will break scale invariance explicitly
\begin{align}
\delta\Gamma[g_{\m\n},\phi]_L=\omega\int d^4 x \sqrt{|g|}\;  \sum_{{\cal O}'} {\cal O}'_L[R_{\m\n\r\s},\phi]
\end{align}

These are precisely the anomalous terms that appear in the Ward identity \eqref{eq:Ward_anomalous} and that will compromise the quantum equivalence of frames
\begin{align}
\bigg\langle 2\mathbf{g}_{\m\n}\frac{\delta \Gamma_J}{\delta \mathbf{g}^{\m\n}}-\bm{\phi}\frac{\delta \Gamma_J}{\delta \bm{\phi}}\bigg\rangle= \sum_{{\cal O}'} {\cal O}'_L[R_{\m\n\r\s},\phi]+\langle \D_\m J^\m\rangle
\end{align}

We could however instead use the dimension dependent transformations \eqref{eq:field_redefinition_n} to come back from the Einstein frame to the Jordan frame. In that case, the counterterms in the latter will contain extra powers of the scalar field inside the integral
\begin{align}
\tilde{\Gamma}[g_{\m\n},\phi]_L=\frac{1}{n-4}\int d^n x \sqrt{|g|}\; \phi^{\frac{4-n}{\lambda_\phi}}\sum_{{\cal O}'} {\cal O}'_L[R_{\m\n\r\s},\phi]+\text{finite}
\end{align}
where $\lambda_\phi=(2-n)/2$ is the scaling dimension of $\phi$ for arbitrary $n$. This represents a regularization scheme that has been used before in the literature\cite{Englert:1976ep,Bezrukov:2010jz,Armillis:2013wya} to relax the tension with equivalence issues like the one presented here, representing a modification of dimensional regularization that seems more suitable for scale invariant theories.

Although in the limit $n\rightarrow 4$ both choices look to be the same, there are some subtle differences. Now the integrand is invariant under scale transformations \emph{in any space-time dimension} and no anomalous term arises. Since such a regularization that preserves scale invariance exists, no anomaly can arise. However, we must have present at all times that dimensional regularization is just a tool and the theory only makes sense once the divergences are substracted and the regularization is removed. If we now renormalize the theory and subtract the pole in $n\rightarrow 4$, we find a new finite contribution of the form
\begin{align}
\tilde{\Gamma}^{\text{(finite)}}[g_{\m\n},\phi]_L=-\frac{1}{\lambda_\phi}\int d^4 x \sqrt{|g|}\; \log(\phi)\sum_{{\cal O}'} {\cal O}'_L[R_{\m\n\r\s},\phi]+\text{other terms}
\end{align}
that will require the introduction of new counterterms which are logarithmic in the field $\phi$ if we want to implement a renormalization scheme that preserves scale invariance.

However, even if these counterterms preserve now scale invariance in the divergent part of the effective action, it is explicitly broken in the finite part. The reason is that the evanescent piece that is left after taking $n\rightarrow 4$ is not scale invariant by itself, it requires the presence of the pole part so that there can be a subtle cancellation. Thus, we find that if we use a minimal subtraction scheme, the finite part will violate the Ward identity precisely as
\begin{align}\label{finite}
\delta \tilde{\Gamma}^{\text{(finite)}}[g_{\m\n},\phi]_L=\omega\int d^4 x \sqrt{|g|}\;  \sum_{{\cal O}'} {\cal O}'_L[R_{\m\n\r\s},\phi]
\end{align} 
which is again the anomalous terms. Therefore, even with this regularization, the anomaly resurges as a finite effect that can modify the theory down to the IR\cite{Alvarez:2015ewa} and it is not clear if there exists some subtraction scheme that can solve this problem. 

Let us finally stress that here we have found that one can choose between the anomaly or the non-polynomial counterterm by choosing if the transformation back to the Jordan frame is analytically continued to arbitrary $n$ or not. However, this is just one way to see how this issue arises. In principle we could compute the counterterms directly in the Jordan frame and there the two options represent two possible choices of regularization scheme. One could stick to standard dimensional regularization, thus finding an anomaly that violates frame equivalence explicitly or, on the other hand, one could use instead the improved regularization scheme, corresponding to adding powers of $\phi$, finding that no anomaly arises but non-standard counterterms are required and the Ward identity is violated anyway. 

Since we are dealing with non-renormalizable theories, these two choices will presumably represent in the Jordan frame two different UV completions. However, they both share the fact that the effective action develops, either through the anomaly or through the finite contribution \eqref{finite}, a non-vanishing S-matrix element which has no counterpart in the Einstein frame, still conveying with our theorem \ref{th:inequivalence}. In this way, equivalence is actually even more compromised, since depending on the choice of transformation, which is isomorphic to the choice of regularization in the Jordan frame\footnote{Indeed, the fact that more new regularizations schemes are available in the Jordan frame when compared to the Einstein frame is another hint towards the problems introduced by the new symmetry.}, we arrive to one theory or another, both presenting physical phenomena which are not contained in the starting theory. At any time, since the theories only exist, in the physical way, once the regularization is removed, we can conclude that different frames lead to different theories.

%%%%%%%%%%%%%%%%%%%%%%%%%%%%%%%%
%%%%%%%%%%%%%%%%%%%%%%%%%%%%%%%%
\section{Conclusions}
We have studied the problem of the quantum equivalence between two different frames of a physical theory, related by a non-linear transformation of the field variables of the form $Q'=Q'(Q)$. When the symmetries of the two frames are realized in a different way, then equivalence could be violated by the effect of anomalous contributions to the Ward-Takahashi identities of some of the symmetries. In such a case, the current associated to the anomalous symmetry represents an operator which is not equivalent in the two frames. In other words, the current does not transform from one frame to the other according to $Q'=Q'(Q)$ and can produce a non-vanishing S-matrix element which is present only in one of the frames.

This is realized in cosmological settings when the inflaton field is driven by a potential which is scale invariant. In the presence of dynamical gravity, there are anomalous corrections to the conservation of the dilatation current that occur only in the Jordan frame, denoting a violation of the equivalence premise with the Einstein frame. Although a regularization scheme, by introducing higher powers of the scalar field, that solves the tension is available, it requires the inclusion of logarithmic counterterms which compromise the structure of the theory down to the IR level, where it breaks scale invariance anyway. Although unpleasant, it seems reasonable to conclude that scalar tensor theories defined in different frames are different physical theories after all and one should be able to perform an experiment to distinguish them.

However, the theories here considered are non-renormalizable and we could question if this effect is just an unpleasant consequence of the former. It would be interesting to find a toy model where this problem appear without the presence of dynamical gravity or non-renormalizable interactions. 

Although we have worked out this issue in a particular model of interest in cosmology, it is not restricted to that setting. The anomalous equivalence could as well arise in the relation of $F(R)$ theories with scalar-tensor ones as well as in any other setting in which a non-linear redefinition of the field variables is done. It would be interesting to study if this situation can be realized in simpler models which do not involve gravity.

Finally, it is worth to comment that even when this issue does not appear, there are still open questions about the equivalence premise, specially in cosmology. When the Universe is de Sitter like, definition of the S-matrix is a subtle issue due to the non-existence of well defined asymptotic states. There, the only trusted observables are equal time correlation functions, for which the question of equivalence is still open. More work is required in order to assert to what extent one can trust in frame equivalence in order to extract physical universal results from the theory.

%%%%%%%%%%%%%%%%%%%%%%%%%%%%%%%%
%%%%%%%%%%%%%%%%%%%%%%%%%%%%%%%%
\section*{Acknowledgements}
I acknowledge discussions with Sergio González-Martín and Carmelo P. Martín during previous collaborations. I also want to thank Enrique Álvarez and Sergey Sibiryakov for useful comments on a first version of this article. My work has been supported by the European Union FP7 ITN INVISIBLES (Marie Curie Actions, PITN-GA-2011-289442) and by the Spanish MINECO Centro de Excelencia Severo Ochoa Programme under grant SEV-2012-0249.
%%%%%%%%%%%%%%%%%%%%%%%%%%%%%%%%
%%%%%%%%%%%%%%%%%%%%%%%%%%%%%%%%

%%%%%%%%%%%%%%%%%%%%%%%%%%%%%%%%
%%%%%%%%%%%%%%%%%%%%%%%%%%%%%%%%
\end{document}